\documentclass{elsart}
\usepackage{graphicx}
\begin{document} 
\begin{frontmatter}
\title{Mass Determination of Groups of Galaxies: Effects of the Cosmological
Constant}

\author{S. Peirani\corauthref{cor1}}
\ead{peirani@obs-nice.fr}
\corauth[cor1]{Corresponding author.}
\author{and}
\author{J.A. de Freitas Pacheco}
\ead{pacheco@obs-nice.fr}
\address{Observatoire de la C\^ote d'Azur, B.P. 4229, F-06304
 Nice Cedex 4, France}
\date{Received date; accepted date}

\begin{abstract}

The spherical infall model first developed by Lema\^itre and Tolman was modified in order to include
the effects of a dark energy term. The resulting velocity-distance relation was
evaluated numerically. This equation, when fitted to actual data, permits the simultaneous
evaluation of the central mass and of the Hubble parameter. Application of this relation to the
Local Group, when the dark energy is modeled by a cosmological constant,
yields a total mass for the M31-Milky Way pair of $(2.5\pm 0.7)\times 10^{12}\,M_{\odot}$, a Hubble parameter
H$_0 = 74\pm 4\,kms^{-1}Mpc^{-1}$ and a 1-D velocity dispersion for the flow of about
39 km$s^{-1}$. The zero-velocity and the marginally bound surfaces of the Local Group are
at about 1.0 and 2.3 Mpc respectively from the center of mass.
A similar analysis for the Virgo cluster yields a mass of $(1.10\pm 0.12)\times 10^{15}\,M_{\odot}$
and H$_0 = 65\pm 9\,kms^{-1}Mpc^{-1}$. The zero-velocity is located at a distance of 8.6$\pm$0.8 Mpc
from the center of the cluster. The predicted peculiar velocity of the Local Group towards Virgo
is about $190\,kms^{-1}$, in agreement with other estimates. Slightly lower masses are derived if
the dark energy is represented by a fluid with an equation of state $P = w\epsilon$ with $w = -2/3$.

\end{abstract}

\end{frontmatter}

\section{Introduction}

New and high quality data on galaxies belonging to nearby groups have improved considerably
estimates of masses and mass-to-light ratios (M/L) of these systems. Searches on the POSS II and
ESO/SERC plates (Karachentseva \& Karachentsev 1998; 2000) as well as ``blind" HI surveys
(Kilborn et al. 2002) lead to the discovery of new dwarf galaxies, increasing substantially
their known population in the local universe. Moreover, in the past years, using HST observations, 
distances to individual members of nearby groups have been derived from magnitudes of the tip of 
the red giant branch by Karachentsev and collaborators (Karachentsev 2005 and references therein), which have
permitted a better membership assignment and a more trustful dynamical analysis. 

Previous estimates of $M/L$ ratios for nearby groups were around $170 M_{\odot}/L_{B,\odot}$
(Huchra \& Geller 1982). However, virial masses derived from the aforementioned data are 
significantly smaller, yielding $M/L$ ratios around 10-30 $M_{\odot}/L_{B,\odot}$. If these
values are correct, the local matter density derived from nearby groups would be only a fraction of the 
global matter density (Karachentsev 2005). However, for several groups the crossing time is comparable
or even greater than the Hubble time and another approach is necessary to evaluate their masses, since
dynamical equilibrium is not yet attained in these cases.

Lynden-Bell (1981) and Sandage (1986) proposed an alternative method to the virial relation in order
to estimate the mass of the Local Group, which can be extended to other systems dominated either by
one or a pair of galaxies. Their analysis is essentially based on the spherical infall model. 
If the motion of bound satellites is supposed to be radial, the resulting parametric equations 
describe a cycloid. Initially, the radius of a given shell embedding a total mass M expands, attains a 
maximum value and then collapses. At maximum, when the turnaround radius R$_0$ is reached, the radial 
velocity with respect to the center of mass is zero. For a given group, if the velocity field close to 
the main body, probed by satellites, allows the determination of R$_0$, then the mass can be 
calculated straightforwardly from the relation

\begin{equation}
M = \frac{\pi^2 R_0^3}{8GT_0^2}
\end{equation}
where T$_0$ is the age of the universe and G is the gravitational constant. 

Data on the angular power spectrum of temperature fluctuations of the cosmic microwave background
radiation derived by WMAP (Spergel et al. 2003) and on the luminosity-distance of type Ia supernovae
(Riess et al. 1998; Perlmutter et al. 1999), lead to the so-called ``concordant" model, e.g., a
flat cosmological model in which $\Omega_m$ = 0.3 and $\Omega_v$ = 0.7. The later density parameter
corresponds to the present contribution of a cosmological constant term or a fluid with negative pressure, 
dubbed ``quintessence" or dark energy. The radial motion leading to the aforementioned $M = M(R_0, T_0)$ relation 
neglects the effect of such a term, which acts as a ``repulsive" force. This 
repulsive force is proportional to the distance and its effect can be neglected if the 
zero-velocity surface is close to the center of mass. Turnaround radii of groups are typically of the 
order of 1 Mpc (Karachentsev 2005), while the characteristic 
radius at which gravitation is comparable to the repulsion force is R$_*$ = 1.1M$_{12}^{1/3}$ Mpc, where 
M$_{12}$ = M/(10$^{12} M_{\odot}$) and the Hubble parameter $H_0$ was taken equal to $70\,\, kms^{-1}Mpc^{-1}$. 
This simple argument suggests that the effect of the cosmological term can not be neglected when deriving 
masses from the $M= M(R_0, T_0)$ relationship. 

In this paper we revisit the velocity-distance relationship when a dark energy term is included
in the dynamical equations and calculate the resulting $M = M(R_0, H_0)$ relation. The presence of 
the dark energy term has been also invoked as a possible explanation for the smoothness of the 
local Hubble flow (Chernin 2001; Teerikorpi, Chernin and Baryshev 2005) being a further reason to 
investigate its effects on the $M= M(R_0, T_0)$ relation. For a cosmological
density parameter associated to the ``vacuum" energy $\Omega_v = 0.7$ (our preferred solution), numerical 
computations indicate that the ``zero-energy" surface, beyond which galaxies will never collapse 
onto the core, is located at about 2.3$R_0$. In order to illustrate our results, some applications 
are made to the Local Group and the Virgo cluster. As we shall see, values of the Hubble parameter 
resulting from fits of the actual data to the velocity-distance relation including a cosmological term,
are in better agreement with recent estimates then those derived from the relation obtained either
by Lyndell-Bell (1981) or Sandage (1986). In Section 2 the relevant equations are 
introduced, in Sections 3 and 4 the results are applied to the Local Group and the Virgo cluster and 
finally, in Section 5 the concluding remarks are given.

\section{The velocity-distance relation}

The evolution of a self-gravitating zero-pressure fluid with spherical symmetry was first considered
by Lema\^itre (1933) and Tolman (1934). The Lema\^itre-Tolman model describes quite well the
dynamics of an extended halo around a bound central core, asymptotically approaching a homogeneous
Friedmann background. In this situation, three main distinct regions can be distinguished: i) the central
core, in which the shell crossing has already occurred, leading to energy exchanges which transform
radial into transverse motion; ii) the zero-velocity surface, boundary which separates infalling
and expanding bound shells and iii) the ``marginally" bound surface (zero total energy), segregating
bound and unbound shells. Density profiles resulting from the Lema\^itre-Tolman model were examined
by Olson \& Silk (1979) and application of this model to the velocity field close to the
Virgo cluster were made by Hoffman et al. (1980), Tully \& Shaya (1984), Teerikorpi et al. (1992) among others.

If displacements of galaxies, here associated to the outer halo shells, develop mainly at low
redhsifts when the formation of the mass concentration around the core is nearly complete (see, for
instance, Peebles 1990), then the equation of motion for a spherical shell of mass $m$, moving radially in 
the gravitational field created by a mass $M$ inside a shell of radius R, including the dark energy term is
\begin{equation}
\frac{d^2R}{dt^2} = -\frac{GM}{R^2} - \frac{(1+3w)}{2}\Omega_vH_0^2R(\frac{a_0}{a})^{3(1+w)}
\end{equation}
where $M = 4\pi\int_0^R r^2\rho_mdr$ and $a$ is the scale parameter (the present value is taken as $a_0$ = 1).
The latter satisfies the Hubble equation
\begin{equation}
(\frac{dlg~a}{dt})^2 = H_0[\Omega_m(\frac{a_0}{a})^3 + \Omega_v(\frac{a_0}{a})^{3(1+w)}]
\end{equation}
Here, the common assumption that the dark energy can be modeled as being a fluid with an equation of state
$P = w\epsilon$ was adopted and in the two equations above, the dependence of the dark energy on the
scale parameter was obtained by solving the energy conservation for such a component.
Eq. (2) is intended to describe the motion of shells in the halo, excluding the central region where 
shell crossing effects have probably already occurred.

Defining the dimensionless variables $y = R/R_0$, $\tau = tH_0$ and $x = a/a_0$, eqs. (2) and (3)
can be rewritten as
\begin{equation}
\frac{d^2y}{d\tau^2} = -\frac{1}{2}[\frac{A}{y^2} + Byx^{-3(1+w)}]
\end{equation}

and

\begin{equation}
\frac{dx}{d\tau} = \sqrt{\frac{\Omega_m}{x} + \frac{\Omega_v}{x^{(1+3w)}}}
\end{equation}

where we have introduced the parameters $A = 2GM/(H_0^2R_0^3)$ and $B = (1+3w)\Omega_v$. These
equations were solved numerically by adopting the following procedure. For a given redshift, the
initial value of the scale parameter is derived as well as the corresponding instant of time from
the Hubble equation. If initially, at high redshifts (here taken around $z \sim 100$), the dark energy 
term is negligible, then using a Taylor expansion of the standard Lema\^itre-Tolman solution, when
the angle parameter $\theta << 1$ (see, for instance, Peebles 1980), the initial values 
of $y$ and its derivative $dy/d\tau$
can be estimated. For a given value of $w$, the parameter $A$ is varied until the condition defining
the zero-velocity surface, e.g., $dy/d\tau$ = 0 at $y$ = 1 is satisfied. 
For the case $w = -2/3$, we have obtained $A = 3.414$ and for the particular case $w = -1$, representing
a cosmological constant, $A = 3.658$. Therefore, the mass inside the zero-velocity radius $R_0$ is

\begin{equation}
M = 1.827 \frac{H_0^2R_0^3}{G} = 4.1\times 10^{12}h^2R_0^3 \,\,\, M_{\odot}
\end{equation}

where $h = H_0/(100~kms^{-1}Mpc^{-1})$ and $R_0$ is in Mpc. In the case $w = -2/3$, the numerical
coefficient is slightly smaller (1.705 instead of 1.827). Comparing with eq. (1), we notice that the 
inclusion of the dark energy term represents, for a given $R_0$, an {\it increase} of 
about 28-38\%  on the mass derived by such a procedure.

Once the parameter $A$ is known, the velocity-distance relation, v = v(R), for different shells at a 
{\it given time} is obtained by varying their energy. Shells with negative energy will expand, halt 
and fall back toward the center, while shells with positive energy expand forever, according to the
aforementioned characterization of regions $ii$ and $iii$. At a given time, there is a 
critical energy $E_c$ which defines the zero-velocity radius. Shells having $E < E_c$ have already 
crossed the turnaround point and are collapsing. Consequently, they have {\it negative} 
velocities. Shells with $E > E_c$ are still expanding and thus have {\it positive} velocities.

For the case $w = -1$, which gives a good representation of actual data, as we shall see in
the next sections, the resulting numerical values are quite well fitted by the relation

\begin{equation}
v(R) = -\frac{0.875H_0}{R^n}(\frac{GM}{H_0^2})^{(n+1)/3} + 1.274H_0R
\end{equation}

with $n = 0.865$. Notice that from the condition $v(R_0) = 0$, eq. (6) is recovered.
It is worth mentioning that solutions with negative energies are possible up to $R = 2.30R_0$,
which defines the marginally bound surface. Beyond this critical radius, only unbound shells 
exist. The solution for $w= -2/3$ differs only marginally and will not be considered
in further analyses.

For comparison, the velocity-distance relationship derived for the case $\Omega_v = 0$ is

\begin{equation}
v(R) = -1.038(\frac{GM}{R})^{1/2} + 1.196H_0R
\end{equation}

which has a slightly flatter dependence on the distance than the precedent equation.

\section{The Local Group}

An immediate application of the velocity-distance relationship is the determination of the 
Local Group mass, concentrated mainly on M31 and the Milky Way, as well as the Hubble parameter itself as
we shall see later. Moreover, the velocity-distance relation gives also an indication of the
dispersion of the peculiar velocities over the Hubble flow. As we have already
mentioned, the local velocity dispersion is known to be quite small (Giraud 1986; Schlegel et al. 1994), a 
fact referred usually as the ``coldness" of the local flow. An investigation of the dynamics of the 
Local Group and its environment by using numerical simulations was performed by Governato et al. (1997), who
concluded that cold dark matter models ($\Omega_m = 1\, or\, \Omega_m = 0.3$) are unable to produce
candidates embedded in regions having ``cold" flows.

As a first step, we have searched to check the accuracy of eqs. (6) and (7) by using 
numerical simulations. We have performed N-body simulations using the adaptive 
particle-particle/particle-mesh $(AP^2M)$
code HYDRA (Couchman et al. 1995), with cosmological parameters h = 0.65,
$\Omega_m$ = 0.3, $\Omega_v$ = 0.7 and $\sigma_8$ = 0.9. The simulation was performed in a periodic
box of side 30$h^{-1}$ Mpc including 256$^3$ particles, corresponding to a mass resolution of
$2.05\times 10^8\,\, M_{\odot}$. The simulation started at z = 49 and ended at the present time.
Halos were initially detected by using a friends-of-friends (FOF) algorithm and, in a second step,
unbound particles were removed by an iterative procedure. Thus, all halos in our catalog are 
gravitationally bound objects. For further details, the reader is referred to 
Peirani et al. (2004).

In our mock halo catalog, several examples of pairs with physical characteristics similar to
the MW-M31 pair can be found. However, most of them have nearby (within 3-5 Mpc) halos of
comparable mass or even higher, which perturb considerably the velocity field.
Here, for illustration purposes only, we consider one case in which the main halos have
masses respectively equal to $9.85\times 10^{11}$ and $6.48\times 10^{11}$ solar masses and are
separated by a distance of 0.69 Mpc, parameters comparable to those of the M31-MW pair.
About 73 subhalos were detected within 3 Mpc of this pair,
but satellites situated at distances larger than 2 Mpc are clearly perturbed by other nearby
structures, since the velocity dispersion increases considerably at those distances. Velocities 
and distances with respect to the center of mass were computed for all these objects.

A simple fitting of these simulated data gives $R_0 = 0.98\pm 0.20$ Mpc and using eq. (6)
with h = 0.65 (adopted in the simulations) a total mass of 
$(1.60\pm 0.32)\times 10^{12}\,\, M_{\odot}$ is obtained. In spite
of the excellent agreement with the actual total mass of the pair, this result is 
somewhat fortuitous since the velocity dispersion of the simulated satellites is rather high. In 
practice, besides random motions, errors in distances or velocities increase considerably the 
uncertainty in the determination of the zero-velocity radius. These difficulties can be 
alleviated by searching the best fit of the v = v(R) relation to data. By varying the mass 
in order to minimize the velocity dispersion and giving a higher weight to the inner 
satellites, one obtains $M = (1.48\pm 0.30) \times 10^{12}\,\, M_{\odot}$. This result seems to be 
more confident to evaluate the uncertainties of the method. It is worth mentioning that quoted
errors are estimates based on the spread of values derived from the fitting procedure and not
formal statistical errors. In figure 1
we show the simulated velocity-distance data for satellites with $R \leq$ 1.8 Mpc and the best
fit solution for eq. (7). The derived 1-D velocity dispersion for this simulated data is 73 kms$^{-1}$,
a value lower than that derived by Governato et al. (1997) but in quite good 
agreement with the $\Lambda$CDM simulations by Macci\`o et al. (2005), who obtained a 
velocity dispersion of about 80 kms$^{-1}$ within a sphere of 3 Mpc radius
(see their Fig. 3). A flat, $\Lambda$
dominated cosmology is able to produce flows on scales of few Mpc around field galaxies ``colder"
than pure dark matter models, but somewhat higher than values derived from actual data, as we 
shall see below.    

\begin{figure*}
\centering
\includegraphics[width=13.cm]{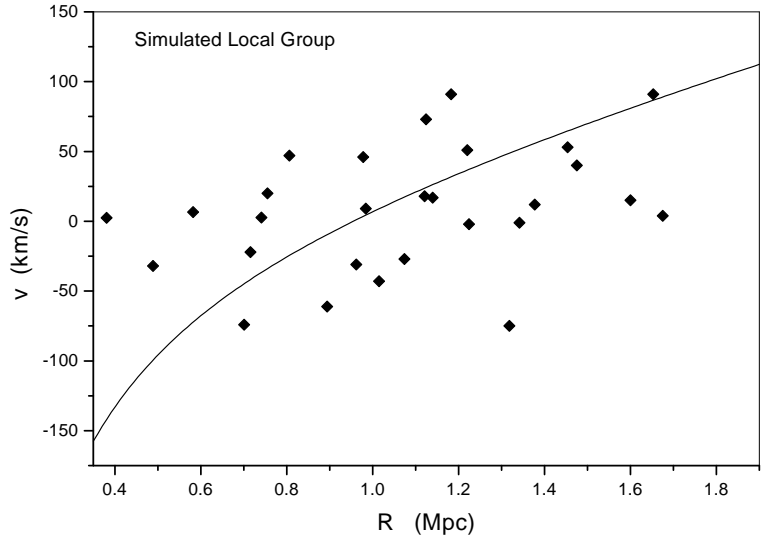}
\caption{Simulated velocity-distance data (diamonds) and best fit to the v=v(R)
relation (solid curve), corresponding to  $M = 1.48\times 10^{12}\, M_{\odot}$ and
$\sigma$ = 73 km/s. The Hubble parameter was held constant, h = 0.65.}
\end{figure*}

Recent data on neighboring galaxies of the Local Group were summarized by Karachentsev et al. (2002), 
who have estimated $R_0 = 0.94\pm 0.10$ and derived from eq. (1) a total mass of 
$1.3\times 10^{12}\,\, M_{\odot}$ for the M31/MW pair.   

Here, eq. (7) was fitted to the data by Karachentsev et al. (2002), but varying now both 
the mass and the Hubble parameter in order to minimize the velocity dispersion. 
We have obtained $h = 0.74\pm 0.04$ and $M = (2.5\pm 0.7)\times 10^{12}\,\, M_{\odot}$, where
the quoted errors are again estimates based on the uncertainties of the fitted parameters. Figure 2 shows 
data points and the velocity-distance relation defined by the previous parameters.
Had we used eq. (8) instead of eq. (7) in the fitting procedure, a
similar result for the mass would have been obtained, but with a {\it higher} Hubble parameter, e.g., 
$h = 0.87\pm 0.05$. We shall return to this point latter.

\begin{figure*}
\centering
\includegraphics[width=13.cm]{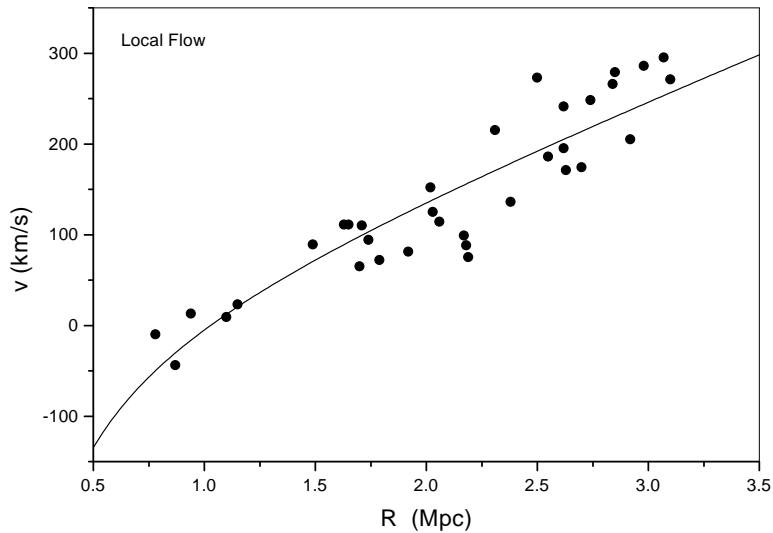}
\caption{Velocity and distance data (circles) for satellites of the M31-MW system
(Karachentsev et al. 2002) and 
the best fit to the v=v(R) relation, corresponding to $M = 2.5\times 10^{12}\, M_{\odot}$ 
and h = 0.74}
\end{figure*}

The zero-velocity radius is located at $1.0\pm 0.1$ Mpc, which is about 10\% higher than the value estimated
by Karachentsev et al. (2002) and the ``marginally" bound surface (zero-energy) is at a distance of
about 2.3 Mpc. From our fit it results a 1-D velocity dispersion of 39 kms$^{-1}$, which
should be compared with the value of 73 km/s found from our simulated data for a similar scale.
Macci\`o et al. (2005) have also revisited the ``coldness" of the Hubble flow and, according to their final 
results, within a sphere of radius 3 Mpc, the expected velocity dispersion is 38 kms$^{-1}$, in rather good
agreement with our figure.

\section{The Virgo Cluster}

Dynamical models for the Virgo cluster based on the Lema\^itre-Tolman model were, for instance, developed
by Hoffman et al. (1980). They have modeled the projected velocity dispersion as a
function of the angular distance and, from comparison with data, derived a mass of $(4.0\pm 1.0)
\times 10^{14}h^{-1}\,\,M_{\odot}$ contained inside a sphere of $6^o$ radius, which corresponds
approximately to the central relaxed core of the cluster. Using the virial relation, Tully \& Shaya
(1984) obtained a mass of $(7.5\pm 1.5)\times 10^{14}\,\,M_{\odot}$ for this central core. More
recently, Fouqu\'e et al. (2001) using the Lema\^itre-Tolman model derived a mass of $1.3\times
10^{15}\,\, M_{\odot}$ inside a radius of $8^o$.

In this section, the derived velocity-distance relation (eq. 7), including effects of the cosmological
constant, will be applied to galaxies outside the inner core of the Virgo cluster. 
 
\begin{figure*}
\centering
\includegraphics[width=13.0cm]{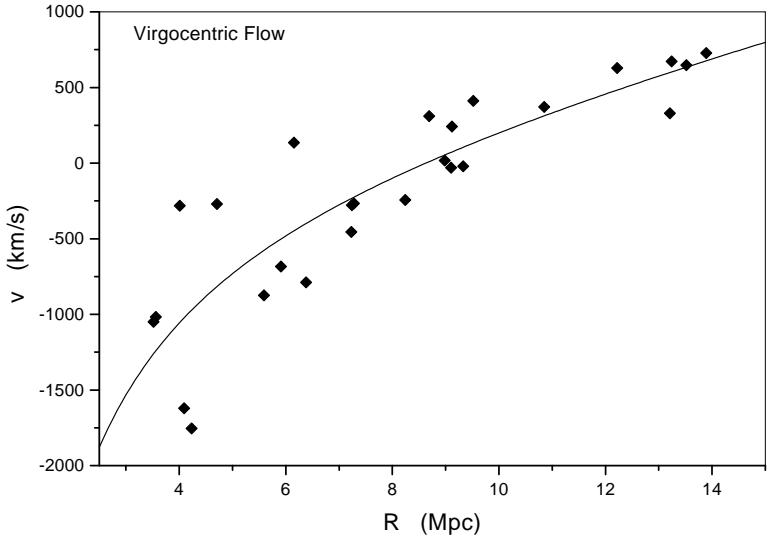}
\caption{Velocity and distance data for galaxies with Virgocentric distances in the
range $3.5 \leq R \leq 15$ Mpc and 
the best fit to the v=v(R) relation, for $M = 1.1\times 10^{15}\, M_{\odot}$ and h = 0.65}
\end{figure*}

Galaxies with Virgocentric distances higher than 1.7 Mpc, corresponding 
approximately to the core radius,
and less than 15 Mpc, were selected from the list by Teerikorpi et al. (1992), constituting
a sub-sample of 27 objects. Distances derived from the Tully-Fisher relation were taken from the 
aforementioned source. Heliocentric velocities
were corrected with respect to the Local Group according to the prescription by Courteau and van den Bergh
(1999). Velocities and distances with respect to Virgo center were calculated by assuming a distance to the cluster
of 16.8 Mpc (Tully \& Shaya 1984) and an observed velocity of 967 km/s (Kraan-Korteweg 1981). 

The best fit of eq. (7) to data gives $h = 0.65\pm 0.09$ and 
$M = (1.10\pm 0.12)\times 10^{15}\,\, M_{\odot}$, corresponding
to a velocity dispersion $\sigma_{1D} = 335$ km/s for the flow. Figure 3 shows the velocity-distance
data for the galaxies of our sample and the theoretical v = v(R) relation computed with the derived
parameters. The higher mass derived in the present analysis confirms some early results based on models 
of the velocity field in the vicinity of the Virgo cluster using the Lema\^itre-Tolman equations, as
those performed by Tully \& Shaya (1998) and Fouqu\'e et al. (2001).
 
The resulting zero-velocity surface is located at $R_0 = 8.6\pm 0.8$ Mpc, which at the assumed distance
corresponds to an angle $\theta_{ZV} = 29^o$. This should be compared with the analyses
by Hoffman et al. (1980), Tully \& Shaya (1984) or Teerikorpi et al. (1992), who estimated
$\theta_{ZV} = 27^o,\, 28^o$\, and\, $25.8^o$ respectively.

Eq. (7), with the above parameters, predicts that at the level of the Local Group the observed velocity
should be 988 km/s, which compares quite well with the observed value. This indicates that
the projection of the peculiar velocity of the Local Group in the Virgo direction is about
190 km/s, compatible with the values found by de Freitas Pacheco (1985), Tamman \& Sandage
(1985) and Federspiel et al. (1998). The marginally bound surface is located
at about 19.8 Mpc, implying that the Local Group is bound to the cluster.

\section{Conclusions}

The contribution of a dark energy term in the mass-energy budget of the universe seems to be well
established at the present time. In this study, the usual velocity-distance relation based
on the Lema\^itre-Tolman model, was revisited in order to include effects due to such a cosmological
term. 

The dynamical equations were solved numerically and the relation $M = M(H_0,R_0)$,
defining the mass inside the zero-velocity surface was recalculated. For a given $R_0$,
the resulting masses are about 35-38\% {\it higher} with respect to the original relation derived 
from the Lema\^itre-Bondi model ($\Omega_v = 0$), if the dark energy is modeled by a fluid with
an equation of state $P/\epsilon = w = -2/3$ or by a cosmological constant respectively.

The resulting v = v(R) relation ($w = -1$ case) was applied to the Local Group and to the Virgo cluster. From the
best fitting procedure, a mass of $(2.5\pm 0.7)\times 10^{12}\,\, M_{\odot}$ was derived for the M31-MW
system, which corresponds to a M/L ratio of about 25 $M_{\odot}/L_{B,\odot}$. The zero-velocity
surface and the ``marginally" bound surface are located at about $1.0\pm 0.1$ Mpc and 
2.3 Mpc respectively. The later defines in fact the boundaries of the Local Group.

More sophisticated analyses of the Local Group based on the action principle were performed by
Peebles (1990). In this method, orbits are reconstructed in order to reproduce the observed
radial velocities, maintaining the action stationary. Peebles (1990) preferred solution implies
a mass of 6.4$\times 10^{12}\, M_{\odot}$ for the Local Group, a factor 2.5 higher than our
result, but a comparison of these results could be meaningless since a different cosmology ($\Omega$ = 0.1)
was adopted in that work. 

The same procedure applied to the Virgo cluster gives a mass of about 
$(1.10\pm0.12)\times 10^{15}\,\, M_{\odot}$
inside a radius of $8.6\pm 0.8$ Mpc. This mass is higher than the virial value (Tully \& Shaya 1984) or
values derived from velocity dispersion profiles computed from models based on the Lema\^itre-Tolman
model (Hoffman et al. 1980). However, recent studies of the velocity field
based on the Lema\^itre-Tolman formulation also lead to masses compatible with our result. Our
velocity-distance relation predicts a peculiar velocity of the Local Group towards the Virgo
cluster of about 190 km/s, compatible with different estimates. The ``marginally" bound surface
encloses the Local Group, which in the future may attain its zero-velocity surface and then 
fall onto the cluster. However this prediction is based on a model in which the Local
Group is considered as an ``isolated" system. In reality
the motion of the Local Group is much more complex, being strongly affected by large mass concentrations
in the direction of Hydra-Centaurus (Great Attractor) and the Shapley supercluster. Some numerical 
simulations predict that within $\sim$ 30 Gyr the Local Group
will get closer to the Virgo center (in comoving coordinates), but then will be pulled 
away(in physical coordinates) due 
to the accelerated expansion of the Universe (Nagamine \& Loeb 2003).    

The introduction of a cosmological constant term modifies the velocity-distance relation in comparison
with that derived from the Lema\^itre-Tolman model. Nevertheless both descriptions of the velocity field near
the Local Group or the Virgo cluster yield masses comparable to within a factor of two. This is probably 
due to the fact that errors still present in distance estimates mask  differences between
both models. However, when searching for a best fit of both models to data, there is a 
substantial difference in the resulting
Hubble parameter. The Lema\^itre-Tolman model requires h in the range 0.87-0.92 in order to fit adequately 
the Local Group and Virgo data respectively, whereas eq.(7) requires h in the range 0.65-0.74, more 
consistent with recent determinations and with the ``concordant" model. In this sense, the inclusion
of the cosmological constant in the v(R) relation seems to improve the representation of actual data. 

\bigskip
\bigskip
\noindent
{\bf Acknowledgement}

\bigskip
\noindent
{We thanks the anonymous referee for his useful comments.
S.\,Peirani acknowledges the University of Nice-Sophia Antipolis for the financial support.}


\small
\begin{thebibliography}{00}

\bibitem{ch01}
Chernin, A., 2001, Physics Usp. 44, 1099
\bibitem{CTP95}
Couchman, H.M.P., Thomas, P.A. and Pierce, F.R., 1995, ApJ, 452, 797
\bibitem{CB99}
Courteau, S. and van den Bergh, S., 1999, AJ 118, 337
\bibitem{FP85}
de Freitas Pacheco, J.A., 1985, AJ 90, 1007
\bibitem{F98}
Federspiel, M., Tammann, G.A. and Sandage, A., 1998, ApJ 495, 115
\bibitem{FSSB01}
Fouqu\'e, P., Solanes, J.M., Sanchis, T. and Balkowski, C., 2001, A\&A, 375, 770
\bibitem{gi86}
Giraud, E., 1986, A\&A 170, 1
\bibitem{go99}
Governato, F., Moore, B., Cen, R., Stadel, J., Lake, G. and Quinn, T., 1997, NewA, 2, 91
\bibitem{HOS80}
Hoffman, G.L., Olson, D.W. and Salpeter, E.E., 1980, ApJ 242, 861
\bibitem{HG82}
Huchra, J.P. and Geller, M.J., 1982, ApJ 257, 423
\bibitem{KK98}
Karachentseva, V.E. and Karachentsev, I.D., 1998, A\&A 127, 409
\bibitem{KK00}
Karachentseva, V.E. and Karachentsev, I.D., 2000, A\&AS 146, 359
\bibitem{Ketal02}
Karachentsev, I.D., Sharina, M.E., Makarov, D.I., Dolphin, A.E., Grebel, E.K., Geisler, D., 
Guhathakurta, P., Hodge, P.W., Karachentseva, V.E., Sarajedini, A. and Seitzer, P., 2002, A\&A 389, 812
\bibitem{K05}
Karachentsev, I.D., 2005, AJ 129, 178
\bibitem{Ki02}
Kilborn, V.A. et al., 2002, AJ 124, 690
\bibitem{KK81}
Kraan-Korteweg, R.C., 1981, A\&A 104, 280
\bibitem{Le33}
Lema\^itre, G., 1933, Ann. Soc. Sci. Bruxelles, A53, 51
\bibitem{LB81}
Lynden-Bell, D., 1981, The Observatory 101, 111
\bibitem{ma05}
Macci\`o, A.V., Governato, F., and Horellou, C., 2005, MNRAS, astro-ph/0512583
\bibitem{NL00}
Nagamine, K. and Loeb, A., 2003, NewA, 8, 439
\bibitem{0S79}
Olson D.W. and Silk J., 1979, ApJ, 233, 395
\bibitem{Pe80}
Peebles, P.J.E., 1980, in The Large-Scale Structure of the Universe, Princeton
Series in Physics, Princeton University Press, New Jersey, p. 80
\bibitem{Pe90}
Peebles, P.J.E., 1990, ApJ, 362, 1
\bibitem{PMP04}
Peirani, S., Mohayaee, R. and de Freitas Pacheco, J.A., 2004, MNRAS, 348, 921
\bibitem{pe99}
Perlmutter, S. et al.,1999, ApJ, 517, 565
\bibitem{Ri98}
Riess, A.G. et al., 1998, AJ, 116, 1009
\bibitem{S86}
Sandage, A., 1986, ApJ, 307, 1
\bibitem{SDS94}
Schlegel, D., Davis, M. and Summers, F.J., 1994, ApJ, 427, 527
\bibitem{Setal03}
Spergel, D.N. et al., 2003, ApJS, 148, 175
\bibitem{TS85}
Tammann, G.A. and Sandage, A., 1985, ApJ, 294, 81
\bibitem{TBGP92}
Teerikorpi, P., Bottinelli, L., Gouguenheim, L. and Paturel, G., 1992, A\&A, 260, 17
\bibitem{tcb05}
Teerikorpi, P., Chernin, A.D. and Baryshev, Yu.V., 2005, A\&A submitted, astro-ph/0506683 
\bibitem{T34}
Tolman, R.C., 1934, Proc. Nat. Acad. Sci. 20, 169
\bibitem{TS84}
Tully, R.B. and Shaya, E.J., 1984, ApJ, 281, 31
\bibitem{TS98}
Tully, R.B. and Shaya, E.J., 1998, in Evolution of the Large Scale Structure, eds.
R.F. Stein and A.G.W. Cameron, ESO, Garching, p. 333
\end{thebibliography}
\end{document}